\documentstyle[fps,twoside,fleqn,espcrc2]{article}
\newcommand{\al}{\alpha}

\newcommand{\gm}{\gamma}

\newcommand{\Sg}{\Sigma}
\newcommand{\dl}{\delta}
\newcommand{\ep}{\varepsilon}

\newcommand{\vep}{\varepsilon}

\newcommand{\kp}{\kappa}
\newcommand{\lm}{\lambda}

\newcommand{\vphibar}{\overline{\varphi}}

\newcommand{\sg}{\sigma}

\newcommand{\ph}{\phi}
\newcommand{\vr}{\varphi}

\newcommand{\ps}{\psi}

\newcommand{\nnn}{\nonumber \\}

\newcommand{\Ps}{\Psi}

\newcommand{\Psb}{\overline{\Ps}}
\newcommand{\psb}{\overline{\ps}}

\newcommand{\mh}{\hat{\mu}}

\newcommand{\chb}{\overline{\chi}}
\newcommand{\hmu}{\hat{\mu}}
\newcommand{\phat}{\hat{p}^2}

\newcommand{\half}{\mbox{{\small $\frac{1}{2}$}} }

\newcommand{\eighth}{\mbox{{\small $\frac{1}{8}$}} }
\newcommand{\Sm}{\sum_{\mu}}
\newcommand{\Tr}{\mbox{Tr}}

\newcommand{\dg}{\dagger}
\newcommand{\ra}{\rightarrow}

\newcommand{\be}{\begin{equation}}
\newcommand{\ee}{\end{equation}}
\newcommand{\bea}{\begin{eqnarray}}
\newcommand{\eea}{\end{eqnarray}}
\newcommand{\eq}{\ref}
\newcommand{\beq}{\begin{equation}}
\newcommand{\eeq}{\end{equation}}
\newcommand{\cc}{\cite}
\newcommand{\lb}{\label}
\newcommand{\aplt}{\mbox{}_{\textstyle \sim}^{\textstyle < }    }

\def \3{\ss}

\newcommand{\AmS}{{\protect\the\textfont2
  A\kern-.1667em\lower.5ex\hbox{M}\kern-.125emS}}

\title{No strong coupling regime in the fermion-Higgs sector of the
standard model \thanks{Presented by W. Bock}}

\author{ Wolfgang Bock
\address{Institute of Theoretical Physics, University of
         Amsterdam, Valckenierstraat 65, 1018 XE Amsterdam,
         The Netherlands},
         Christoph Frick
\address{HLRZ c/o KFA J\"ulich, P.O. Box 1913, 5170 J\"ulich,
         Germany and \\
         Institute of Theoretical Physics E, RWTH
         Aachen, Sommerfeldstr., 5100 Aachen, Germany},
         Jan Smit$^{\;\; {\rm a}}$ and
         Jeroen C. Vink
\address{University of California, San Diego,
         Department of Physics-0319,
         La Jolla, CA 92093-0319, USA}
        }

\begin{document}
\begin{abstract}
We present results for the renormalized quartic
self-coupling $\lm_R$ and the
renormalized Yukawa coupling $y_R$ in a fermion-Higgs model
with two SU(2) doublets, indicating that these couplings are not
very strong.
\end{abstract}

\maketitle
\section{INTRODUCTION}
It is an important issue to investigate within a non-perturbative
regularization scheme whether the quartic self-coupling and Yukawa coupling
of the fermion-Higgs sector of the Standard model (StM) remain
relatively small  when
increasing the bare couplings to very large values. For this
it is desirable to construct a lattice fermion-Higgs model
with a realistic fermion content.
A naive transcription of the continuum lagrangian with
one SU(2) doublet leads to the large  number of
16 doublets on the lattice because of the species doubling phenomenon.
There are two proposals  which allow
to reduce this large number of mass-degenerate SU(2) doublets to one:
The mirror fermion model \cc{Montvay} and the reduced staggered fermion model
\cc{Sm88}.
The mirror fermion model is discussed in ref.~\cc{Li}.
In this contribution we use the reduced staggered formalism \cc{letter,paper}.
The basic idea here is to couple the two reduced staggered flavors
to the Higgs field. For this purpose
it is convenient to introduce the $4\times4$ matrix fields \cc{Sm88}
\bea
\Psi_x&=&\eighth \sum_{b} \gm^{x+b}
\half (1-\ep_{x+b}) \chi_{x+b} \;\;,\;\;\; \nnn
\Psb_x&=&\eighth \sum_{b} (\gm^{x+b})^{\dg}
\half (1+\ep_{x+b})  \chi_{x+b} \;.\lb{SDTR}
\eea
with $\gm^{x}=\gm_1^{x_1} \cdots \gm_4^{x_4}$, $\ep_x=(-1)^{x_1+
x_2+x_3+x_4}$ and the sum running over the corners of a hypercube,
$b_{\mu}=0,1$. The one-component staggered fermion field $\chi$ is a real
Grassmann variable. In contrast to usual staggered fermions
(s.~eq.~(1) of ref.~\cc{Jeroen}) we
inserted here the factors $\half (1-\ep_x)$
and $\half (1 + \ep_x)$ which are used in the reduced
staggered formalism to project the usual staggered
fields $\chi$ and $\chb$ to the odd and even sites of the hypercubic lattice.
The form (\eq{SDTR}) implies the following structure for the matrix fields
\be
\Psb =    \left( \begin{array}{cc}
                     \psb_L &   0    \\
                        0   & \psb_R \end{array} \right),\;
      \Psi =    \left( \begin{array}{cc}
                        0    & \ps_R \\
                       \ps_L &    0    \end{array} \right) \;,
                       \lb{MATRIX}
\ee
where $\psb_L$, $\psb_R$, $\ps_L$ and $\ps_R$ are $2 \times 2$
matrices. The row (column) indices of the $\ps_L$ and $\ps_R$
fields act as Weyl-spinor (flavor) labels, and
vice versa for $\psb_L$ and $\psb_R$.
When using the relation (\eq{MATRIX}) and introducing the
$4\times 4$ matrix for the O(4) Higgs field $\phi$,
\be
      \Phi =    \left( \begin{array}{cc}
                      0      & \ph      \\
                   \ph^{\dg} &  0      \end{array} \right) =
            - \Sm \vr_{\mu}\gm_{\mu},
                 \lb{DEFPHI}
\ee
one can show that the action
\bea
S_F&=&-\sum_x [\Sm \half \Tr
          (\Psb_x \gm_{\mu}\Psi_{x+\mh} -
            \Psb_{x+\mh} \gm_{\mu}\Psi_{x} ) \nnn
 &&\mbox{} +   y \Tr (  \Psb_x \Psi_x \Phi_x^T)   ] \lb{SFM}
\eea
reduces in the classical continuum limit to the action of the fermion-Higgs
sector of the StM with one mass-degenerate isospin doublet.
\begin{figure}[t]
\centerline{
\fpsxsize=5.5cm
\fpsbox{Aphasd.ps}
}
\vspace*{-0.9cm}
\hspace*{0.6cm}
\centerline{
\fpsxsize=7.0cm
\fpsbox{ALindn.ps}
}
\vspace*{-1.4cm}
\caption{Upper figure: Phase diagram at $\lm=\infty$ with $N_D=2$.
Lower figure: $m_{\sg}/v_R$ as a function of $m_F/v_R$.
}
\label{fig:1}
\end{figure}
After inserting (\eq{SDTR}) and (\eq{DEFPHI}) into
eq.~(\eq{SFM})
the final form of the fermionic action in terms
of the $\chi$ fields reads
\bea
 S_F = -\half  \sum_{x \mu} \chi_x \chi_{x+\hmu}
       ( \eta_{\mu x} + y \ep_x \zeta_{\mu x} \vphibar_{\mu x} )
       \;, \lb{SCHI}
\eea
where $\vphibar_{\mu  x} = \frac{1}{16} \sum_b \varphi_{\mu, x-b}$
is the average of the scalar field over a lattice hypercube and
$\eta_{\mu x}=(-1)^{x_1+\cdots +x_{\mu-1}}$,
$\zeta_{\mu x}=(-1)^{x_{\mu+1}+\cdots +x_4}$ are the usual staggered
sign factors. The total form of the action is given by
$S=S_F+S_H$, where
$S_H= \sum_{x}[2\kp \sum_{\mu} \varphi_{\al x} \varphi_{\al,x+\hmu} -
\varphi_{\al x} \varphi_{\al x} - \lm
(\varphi_{\al x} \varphi_{\al x}-1)^2]$ is the pure
scalar field action. The action $S$ is invariant under the so-called
staggered fermion (SF) symmetry group which includes shifts by one lattice
distance, 90$^o$ rotations, lattice parity and the global U(1) symmetry,
$\chi_x \ra e^{i \alpha \ep_x}  \chi_x$. This invariance of $S$
ensures the staggered flavor interpretation in the scaling region.

$S$ is, however, not invariant
under the full O(4) flavor group:
There are two operators with dimension four which are generated by
the quantum fluctuations and which are invariant under
the SF symmetry group, but break O(4),
\be
O^{(1)}=  \sum_{x \mu} \varphi_{\mu  x}^4 , \;
O^{(2)}=  \frac{1}{2} \sum_{x \mu} (\varphi_{\mu , x+\hmu} -\varphi_{\mu  x})^2
. \lb{C2}
\ee
In order to recover the full O(4) symmetry one has in principle to add
these operators as counterterms to the
action $S \ra S + \vep_0 O^{(1)} + \dl_0 O^{(2)}$
and tune the coefficients $\vep_0$ and $\dl_0$ as a function
of the bare parameters such that the O(4) invariance
gets restored in the scaling region. Here we shall not add these counterterms
to the action. However, we will show in the next section that the effect of the
symmetry breaking is small in the parameter region of interest.

Since we are interested in the largest possible renormalized couplings
we have fixed in the numerical
simulation $\lm=\infty$.
For the use of the Hybrid Monte Carlo algorithm
it is necessary to use two mass-degenerate doublets, $N_D=2$.
The $\kp$-$y$ phase diagram is
shown in fig.~1.
There are four different phases, a paramagnetic (PM), a broken
or ferromagnetic (FM), an antiferromagnetic (AM) and a ferrimagnetic (FI)
phase. The various symbols mark the points in the FM phase where we
carried out numerical simulations on lattices ranging in size
from $6^3 24$ to $16^3 24$.
\section{O(4) SYMMETRY BREAKING}
To estimate the amount of O(4) symmetry breaking we have computed the
one- and two-point functions in the FM phase using renormalized
perturbation theory. We can decompose
the scalar field in the FM phase in a Higgs mode, $\sg$,
and three Goldstone modes, $\pi^a$, $a=1,2,3$,
according to $ \vr_{R\mu}=(v_R+\sg_R) e^4_{\mu}+ \pi^a_R e^a_{\mu} $,
where $\{e^{\al}_{\mu}\}$ form an orthogonal set of O(4) unit
vectors, which is arbitrary when neglecting fermion loop
effects. This arbitrariness is removed after taking into account
the one fermion loop contribution to the
vacuum expectation value. One can show \cc{letter,paper} that
the direction of spontaneous symmetry breaking
$e_{\mu}^4$ is compatible with the one fermion loop correction only if the
$e_{\mu}^4$ = $(\pm 1,0,0,0)$, \ldots,
$(0,0,0,\pm 1)$, $(\pm 1,\pm 1,0,0)/\sqrt{2}$,
\ldots, $(\pm 1,\pm 1,\pm 1,\pm 1)/2$.
{}From a computation of the one fermion loop effective potential
for small $y_R$ one finds that only the directions
$e_{\mu}^4 = (\pm 1,\pm 1,\pm 1,\pm 1)/2$
correspond to local minima,
the others are saddle points or local maxima.

The calculation of the one fermion loop contribution to the
self-energy $\Sg_{\mu\nu}(p)$ shows that both terms in eq.~(\eq{C2})
are generated by quantum fluctuations and should therefore be included
in the tree level effective action
with $\dl_0, \vep_0 \ra \dl_R, \vep_R$.
Explicit expressions for $\Sg_{\mu\nu}(p)$ and the coefficients
$\vep_R$ and $\dl_R$ are derived in
ref.~\cc{paper} in renormalized perturbation theory,
$\vep_R=f_{\vep}(m_F) N_D y_R^4$,
$\dl_R = f_{\dl}(m_F) N_D y_R^2$. The lattice
integrals $f_{\vep}(m_F)$ and $f_{\dl}(m_F)$, whose explicit form
is given in ref.~\cc{paper}, can be computed numerically,
$f_{\vep}(m_F)=0.0054,0.0043$, $f_{\dl}(m_F)=0.026,0.017$ for $m_F=0,0.5$.
{}From the one loop result for the renormalized propagator we can
read off the following estimate for the Goldstone mass,
\be
m_{\pi}^2=\frac{2 \vep_R v_R^2}
                  {1+\dl_R/4}.  \lb{M}
\ee
\begin{figure}[t]
\centerline{
\fpsxsize=6.5cm
\fpsbox{Ampi.ps}
}
\vspace*{-1.6cm}
\caption{Goldstone mass as a function of $y$.
}
\label{fig:2}
\end{figure}
The $\vep_R$-term in the effective action gives
rise to the non-zero value of the Goldstone mass.
Fig.~2 shows that the numerical values (squares) for $m_{\pi}$
are very small in the scaling region with
$y\approx 3.6-4.0$. Moreover the numerical results for $m_{\pi}$
are in good agreement with the analytic prediction (\eq{M}) (diamonds)
after inserting the measured values for $y_R$ and $m_F$.
This motivates us to take also the corrections for the renormalized
field expectation value and Higgs mass
in eq.~(\eq{M}) seriously and to define corrected couplings,
$y_R^{\prime}=y_R (1+\frac{\dl_R}{4})^{1/2}$,
and $\lm_R^{\prime} =
\frac{m_{\sg}^2}{2v_R^2}(1-\frac{m_{\pi}^2}{m_{\sg}^2})
(1+\frac{\dl_R}{4})^2$.
A measure for the O(4) symmetry breaking corrections is given by
the ratios $R_y=(y_R-y_R^{\prime})/y_R^{\prime}$ and
$R_{\lm}=(\sqrt{2\lm_R}-\sqrt{2\lm_R^{\prime}})/ \sqrt{2\lm_R^{\prime}}$.
A numerical
calculation of these ratios gives $|R_y| <5\%$ and $|R_{\lm}| < 7\%$,
in a parameter region with $m_F<0.5$ and $m_{\sg} <0.7$, which shows
that the symmetry breaking effects are small.
\section{RESULTS OF THE SIMULATION}
Since the effect of the symmetry breaking is small we have computed
the renormalized couplings from the usual
tree level relations $y_R=m_F/v_R$ and $\lm_R=m_{\sg}^2/2v_R^2$,
where the renormalized field expectation value is defined as
$v_R=v/\sqrt{ Z_{\pi} }$. Here $v$ is the unrenormalized scalar
field expectation value and $Z_{\pi}$
the wave-function renormalization constant of the Goldstone propagator.
For the determination of the quantities $m_F$, $m_{\sg}$ and $Z_{\pi}$
we have measured the fermion,
$\sg$ particle and Goldstone propagators
in momentum space. The fermion propagator could be well described for all
momenta by a one pole Ansatz, which is characteristic for
weakly interacting fermions. For the Goldstone propagator we have
displayed a typical example in fig.~3. The inverse
propagator $G^{-1}_{\pi}$ is plotted here as a function of $\phat$,
the square of the lattice momentum. The numerical data (crosses) exhibit
a significant curvature at small $\phat$ which can be described by
the one fermion loop
contribution to the self-energy. This non-linear $\phat$ dependence
can be parametrized by the  Ansatz
$G_{\sg,\pi}^{-1}(p)=(\phat + m_{\sg,\pi}^2+\Sigma_{sub}(p))/Z_{\sg,\pi}$,
where $\Sg_{sub}$ is the subtracted one fermion loop self-energy.
The circles in fig.~3
were obtained by fitting this Ansatz to the numerical data.
\begin{figure}[t]
\hspace{0.1cm}
\centerline{
\fpsxsize=7.0cm
\fpsbox{Api038.ps}
}
\vspace{-1.6cm}
\caption{$G^{-1}_{\pi}(p)$ as a function of $\phat$.
}
\label{fig:3}
\end{figure}
The fact that
the one loop Ansatz is sufficient to describe the numerical results
perfectly over a large momentum interval  indicates already
that the renormalized couplings are small.
This fitting method allows us to determine
$m_{\sg}$ and $Z_{\pi}$ accurately, also on small volumes.

As a next step we have to extrapolate the finite volume results
for $y_R$ and $\lm_R$ to the infinite volume. We carried out
simulations on lattices of size $L^3 24$ with $L$ ranging from $6$ to
$16$. If the spectrum contains massless Goldstone bosons, this gives
rise to a $1/L^2$ dependence of the
finite volume quantities. Since the Goldstone particles
are massive in our model we expect deviations from the
linear $1/L^2$ dependence when the volume increases beyond the
Goldstone correlation length, $L > O(1/m_{\pi})$. The fact that we did not
observe significant deviations
gives further evidence that the symmetry breaking effects are small.

In the lower graph of fig.~1 we display the infinite volume results
for the ratios $m_{\sg}/v_R=\sqrt{2\lm_R}$ and $m_F/v_R=y_R$. The symbols
in the upper and lower diagrams of fig.~1 match,
so that one can see
where in the phase diagram the results for the ratios have been
obtained. It can be seen that the numerical values for neither
ratio change when lowering $\kp$ beyond
$\kp=0$, while keeping the cut-off roughly constant.
The $v_R$ values of these points vary from  $0.08$ to $0.27$.
The arrows in fig.~1 mark the
tree level unitarity bounds for $\lm_R$
and $y_R$. The graph shows that
the points obtained in the regions (II) and (III)
of the phase diagram
(see fig.~1) are still very close to these values,
which indicates that
the renormalized couplings are not very strong.
The solid line
encloses the allowed regions obtained
by integrating the one loop $\beta$ functions from infinite couplings
at the cut-off downward to the renormalization scale.
The cut-off was adjusted such that the agreement with
the numerical data is best. It is remarkable that the
shape is in reasonable
agreement with our data.
Fig.~1 shows that the Yukawa interaction gives a slight
increase in $\lm_R$.
{}From fig.~1 we can read off an upper bound
for $m_{\sg}/v_R$ and $m_F/v_R$:
For $m_{\sg} \aplt 0.7/a$, we find $m_{\sg}/v_R \aplt 4$
and $m_F/v_R \aplt 2.6$. From experience in the O(4) model
with various re\-gularizations, we expect that
these numbers for the upper bounds may
be stretched by perhaps 20-30\%.

All in all we conclude that the renormalized
quartic and Yukawa couplings are in accordance with
triviality and that they cannot be strong, unless
the cut-off is unacceptably low.\\

The numerical calculations were performed on the CRAY Y-MP4/464
at SARA, Amsterdam, on the S600 at RWTH Aachen
and on the CRAY Y-MP/832 at HLRZ J\"ulich.
This research was supported by the ``Stichting voor
Fun\-da\-men\-teel On\-der\-zoek der Materie (FOM)''
and by the ``Stichting Nationale Computer Faciliteiten (NCF)''.
%
%

\end{document}